%% file: Luttinger.tex
\documentclass[twocolumn,
preprintnumbers,amsmath,amssymb]{revtex4}
\usepackage{graphicx}
\usepackage{dcolumn}
\usepackage{bm}


\begin{document}

\preprint{preprint}
\title{Study of the charge correlation function in one-dimensional Hubbard heterostructures}
\author{Y. Arredondo}
\author{H. Monien}
\affiliation{Physikalisches Institut, Universit\"at Bonn,
             Nussallee 12, 53115 Bonn, Germany}
           
\begin{abstract}
  We study inhomogeneous one-dimensional Hubbard systems using the density
  matrix renormalization group method. Different heterostructures are
  investigated whose configuration is modeled varying parameters like the
  on-site Coulomb potential and introducing local confining potentials. We
  investigate their Luttinger liquid properties through the parameter
  $K_\rho$, which characterizes the decay of the density-density correlation
  function at large distances. Our main goal is the investigation of possible
  realization of {\it engineered materials} and the ability to manipulate
  physical properties by choosing an appropriate spatial and/or chemical
  modulation.
\end{abstract}

\maketitle

\section{\label{sec:intro}Introduction}

A key aspect of materials research is to find parameters to tune the physical
characteristics of the system like conductivity and other desired properties.
In the last decades there has been enormous progress in the generation of
nanoscopic quasi-one-dimensional systems, e.~g., carbon nanotubes
\cite{A02_Iijima,A02_Ishii}, semiconducting quantum wires
\cite{01_Tarucha,A04_Auslaender} and organic molecules \cite{01_Bechgaard}; as
well as an intense study of their transport properties
\cite{25A_Auslaender,22_Bockrath,20_Degiorgi} such as superconductivity
\cite{19A_Jerome} and quantum Hall edge states \cite{19_Milliken,19B_Chang} .
While the properties of homogeneous one dimensional systems (even with
disorder) are relatively well understood, very little is known about the
properties of strongly interacting inhomogeneous systems. Despite of the large
effort in the study of heterostructures and quantum dots
\cite{01_Hallberg,01_Riera,01_Dagotto,25_Ejima,01_LLS}, there are still open
questions which become relevant in modeling the transport through molecules,
where the electrons interact strongly due to the reduced dimension.  In
addition, its chemistry induces potential barriers which alter the transport
properties drastically.  Technically it is very important to know how to
control the transport and equilibrium properties. In this paper we present a
detailed investigation of correlation effects in an inhomogeneous
one-dimensional system including potential barriers.

The strong electron correlations, inherent to the low-dimensional structure,
and the large quantum fluctuations induce new and interesting quantum phases.
The relevant degrees of freedom are no longer the single particle electronic
states but the collective spin and charge density waves. The low-energy
electronic single particle excitations possesses vanishing spectral weight at
the Fermi surface. The physics of such systems, in the homogeneous low-energy
regime, is well described by the Tomonaga-Luttinger liquid (TLL) model
\cite{01_Luttinger, 01_Tomonaga} introduced by Haldane \cite{01_Haldane}.
Within this model, it is found that all correlation functions exhibit a
power-law decay with the distance, which is specified only by the parameter
$K_\rho$, known as the Tomonaga-Luttinger (TL) parameter.

For inhomogeneous structures the high-energy physics is determined by the
underlying chemistry which, in the atomic scale, introduces Coulomb
correlations and local potentials. On the other hand, at large length scales,
the physics has to be described by the TLL model. In order to establish a
connection between the low-energy TLL and the quasi-one-dimensional systems
synthesized in the laboratory, we investigate the density-density correlation
functions in the asymptotic region (i.~e. for well separated positions $ x $
and $ x'$). Position dependent on-site Coulomb interaction $U(x)$ and a local
potential $V(x)$ are used to model the changes in the local chemistry of the
heterostructures. This defines regions which, for slowly varying potentials,
can be separately considered as homogeneous. We wish to study how the TL
parameter changes close to the crossover regions. We expect to find a
description of it in terms of $U(x)$ and the local density $n(x)$.

The paper is organized as follows: In Sec.~\ref{sec:model} the composition of
the investigated heterostructures is described and we plot our expectations in
terms of the coupling parameters. In Sec.~\ref{sec:method} we briefly recall
the approximate results in the low-energy regime for correlation functions in
the homogeneous case, and we describe the numerical procedure, the DMRG
method, used to study the one-dimensional heterostructures.  The results are
presented and discussed in Sec.~\ref{sec:results}. Finally we state our
conclusions.

\section{\label{sec:model}Hubbard Heterostructures}
\begin{figure}[h]
  \includegraphics[scale=0.45]{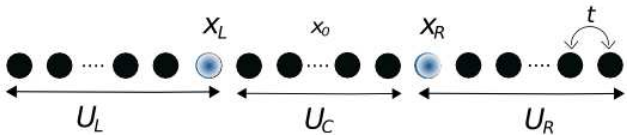}
  \caption{\label{fig:HSL} General arrangement of a Hubbard heterostructure. 
    The measurements for $\langle n(x)n(x_0) \rangle$ were carried out from
    the middle point $x_0=120$.}
\end{figure}

The Hubbard heterostructures we investigate are chains with a length of $L$
sites and on-site Coulomb interaction $U$, which switches between two
different values. In our case it can be visualized as a valley around the
middle of the chain with sharp edges at the sites labeled $x_L$ and $x_R$ .
We will refer to this system as {\it Heterostructure I}. We expect that the
slight discontinuity in the charge distribution, caused by this form of
interaction, will not strongly affect the correlation between the adjacent
regions and will make it possible to find a TLL behavior, even in the region
after the change in the $U$ interaction. On a second heterostructure (called
from here on {\it Heterostructure II}), in addition to the Coulomb interaction
described, two potential walls are introduced through the confining potential
$V$($\gg U$).  Because of the sharp discontinuity in the charge distribution,
we do not expect to find a TLL extending beyond the point of the scattering
potential, however it might still be possible to approximate the TLL in the
different subchains, since in each of them we expect to find a homogeneous
particle distribution.  Fig.~\ref{fig:HSL} shows the layout of the
heterostructures.

Our starting point is an inhomogeneous form of the Hubbard Hamiltonian:
\begin{equation}
  H = -t \sum_{i,\sigma}c_{i,\sigma}^{\dagger}c_{i+1,\sigma} +
  \sum_{i}U_in_{i\uparrow}n_{i\downarrow} +
  \sum_{i,\sigma}V_in_{i\sigma} 
  \label{Hubbard}
\end{equation}

\noindent where $c_{i,\sigma}^{\dagger}$ ($c_{i,\sigma}$) is the creation (annihilation)
operator with spin $\sigma$ ($=\uparrow,\downarrow$) on the site $i$ and
$n_{i\sigma}=c_{i,\sigma}^{\dagger}c_{i,\sigma}$ is the electron number
operator. $t=1$ is the nearest neighbor hopping matrix, which we choose to set
the energy scale. The Hamiltonian in Eq.~(\ref{Hubbard}) incorporates the
different systems we want to study and will allow us to find out if such
systems resemble a TLL and, in that case, also to determine the $K_{\rho}$
parameter from the density-density correlation function.

The sites $x_L$ and $x_R$ divide the whole system into three homogeneous
subchains $U_L, U_C$ and $U_R$, raising two questions: first, how the charge
correlation function behaves in the whole system and second, whether the known
results for the homogeneous regime can be recovered within the subchains.

\section{\label{sec:method}Approximate description of one-dimensional systems 
in the low-lying energy sector}

The low-lying energy, long-distance physics of one-dimensional fermionic
systems is described by bosonic collective excitations. This bosonization
technique yields an exact solution for the TL model, whose complete
description depends solely on the charge and spin velocities and the TL
parameter $K_{\rho}$. In the next subsections we first briefly recall the
known results \cite{01_Schulz,Giamarchi_Schulz} for the density correlation
function in the case of homogeneous systems and explain in detail the
numerical method used to measure the correlation functions in the
inhomogeneous systems.
\subsection{Homogeneous regime}
In a homogeneous TLL, $K_{\rho}$ determines the long-distance decay behavior
of all the correlation functions. In the absence of external magnetic field or
spin anisotropic interactions, the charge correlation function is given by

\begin{eqnarray}
  \small{\langle n(x)n(0)\rangle=\frac{K_{\rho}}{(\pi x)^2} +
    \frac{A_1\cos(2k_{\rm{F}}x)}{x^{1+K_{\rho}}}\mathrm{ln(x)}^{-\frac{3}{2}}} \nonumber\\
  \small{+\frac{A_2\cos(4k_{\rm{F}}x)}{x^{-4K_{\rho}}} + ...}
  \label{correlate}
\end{eqnarray}

Even though the constant coefficients $A_1, A_2$, and $B_1$ depend on the
model, the algebraic decay is characterized only by $K_{\rho}$. Of special
physical interest are the charge density waves with wave vectors $2k_{\rm{F}}$
and $4k_{\rm{F}}$.  While the $2k_{\rm{F}}$ mode dominates over the
$4k_{\rm{F}}$ for $K_{\rho}\ge\frac{1}{3}$, for sufficiently large values of
the on-site Coulomb interaction $U$, the $4k_{\rm{F}}$ charge mode dominates
over the $2k_{\rm{F}}$ mode.

\begin{figure}[h]
  \includegraphics[scale=0.60]{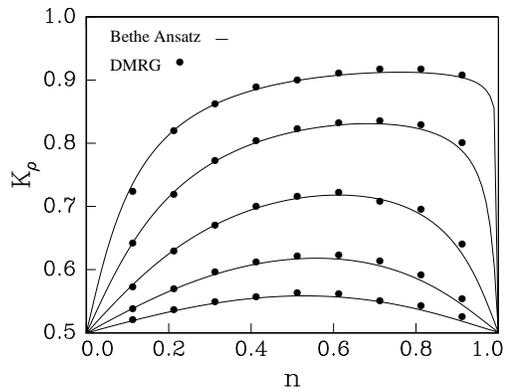}
  \caption{\label{fig:DMRG} Tomonaga-Luttinger parameter values for the
    Hubbard lattice compared to our numerical evaluation of $K_\rho$ (dots).
    $U=1.0,2.0,4.0,8.0,16.0$ from top to bottom.}
\end{figure}

As a test for our numerics, we considered the case of a homogeneous chain for
which we confirmed the results obtained from the Bethe Ansatz \cite{01_Schulz,
  25a_Ejima} for the correlation functions. In Fig.~\ref{fig:DMRG} we show our
results for several values of $U$ obtained with a homogeneous chain of length
$L=240$ sites. We will use this form of the density correlation function to
analyze the low-energy behavior of the Hubbard heterostructures.

\subsection{Inhomogeneous regime\\Numerical study}
The measurement of observables, which include ground state energies and
correlation functions, is carried out using the density matrix renormalization
group (DMRG) \cite{Whi92, Whi93, Peschel}, a method whose roots go back to the
numerical renormalization group formulated by Wilson \cite{Wilson}. The DMRG
is an efficient numerical method developed to overcome the intrinsic
difficulties of low-dimensional strongly interacting systems.

The DMRG provides two algorithms to handle an otherwise
exponentially-increasing Hilbert space of a many-body system. Both
implementations, finite-size and infinite-size DMRG base, as in Wilson's
renormalization group, on a blocking treatment of a lattice system in
real-space, whose basis of the corresponding Hilbert space is decimated under
a certain criterion. In the renormalization group procedure, the decimation of
the system's basis is done by selecting $m$ states with the lowest energy
eigenvalues to obtain the ground state of a system.  This proved to be a
reliable method to solve systems, such as the Kondo problem, for which the
coupling between successive sites decreases exponentially. Thus, it was
plausible to ignore the connections between neighboring blocks, setting to $0$
the wave function at the sites outside of the block of interest. This lead to
inaccuracies when studying systems such as the Hubbard model, where there is
no intrinsic separation of the energy scales. To solve this, White proposed
other criteria to handle both the boundary conditions when adding a new site
to the system as well as the selection of states to best represent it.

The DMRG method considers the system to be connected to a bath, which is a
second block, forming in total a {\it superblock}. The interactions between
the system and the bath set the boundary conditions at the edge sites of the
system as if it would be part of a larger system. In this way, the procedure
becomes more accurate as the system gets larger. The wave function in the
superblock has the form $|\Psi\rangle = \sum_{i,j} \Psi_{ij} |i\rangle \otimes
|j\rangle $, where $i$ are the states on the system and $j$ are those on the
bath. From this, the reduced density matrix of the system is $\rho_{ii'}=
\sum_{j}\Psi_{ij}\Psi_{i'j}$. The crucial point is that the density matrix
contains all the information needed to calculate any property of the system
and so, the state of the system can be optimally represented by keeping the
$m$ most probable states given from the density matrix of the system.

We use the finite-size DMRG algorithm. This consist of the following steps:
After growing our system up to a fixed size $L$, by means of the infinite-size
DMRG, the basis of this final system is optimized to best represent the
desired target state, like the ground state, by sweeping through the system
repeatedly.  A {\it sweep} over the system is an iterative process which
starts with a small block on the right extreme of the chain. This is grown to
in the left direction by adding a site to the right block and connecting it to
a bath or environment on the left side. The environment information was
collected from the infinite-size algorithm. The total size of the system is
always kept constant.  As soon as the decreasing size of the left block
reaches a single site the procedure is stopped. We save the information of the
right blocks and can use it now to start a similar procedure with a block on
the left side of the chain being grown in the right direction.  This procedure
is repeated until convergence is reached.

With each step, the chain grows one site in the current direction, and the
basis of the new system must be truncated to keep the Hilbert space
manageable. All the necessary operators are transformed and stored every time
this happens. With every step, the choice of states in the truncation of the
basis becomes a better representation of the system. This leads to an optimal
truncated basis for representing the target state on the finite system. After
convergence was reached, we can proceed to measure other observables.

The numerical error caused by truncation of the original basis can be measured
through the weight of the states that were discarded in a DMRG step.  Our
systems, with $L=240$ sites under open boundary conditions, were investigated
keeping $m=256$ density-matrix states, rendering a maximum truncation error of
approx. $10^{-6}$.

\subsection*{Density-density correlation function}
An operator $A$, acting either on the left or on the right block, can be
written in the basis of the specific block as $\langle\Psi|A|\Psi\rangle$.  In
the case of correlation functions like $\langle\Psi|AB|\Psi\rangle$, handling
operators requires some extra attention. The operators $A$ and $B$ can operate
either on equal or on different blocks. The last case may lead to errors in
the calculation of the expectation value of the product $AB$, since each
operator is separately written in its corresponding basis. The way to proceed
is to build the exact operator $C=AB$, in a full basis from the beginning and
transform it as is done for the rest of the operators.
 
We calculated the TL parameter $K_{\rho}$ by measuring the correlation
function between the sites $x$ and $x_0$: $C_{x}=\langle
n(x)n(x_0)\rangle-\langle n(x)\rangle\langle n(x_0)\rangle$, where the static
expectation values were subtracted.  To reduce the effect of the local density
oscillations, we take the average over pairs of correlation functions for
neighboring sites calculating $C(r)=(C_{x}+C_{x+1})/2$, with $r=|x-x_0|$ and
$x_0$ in the middle point of the chain. Due to the symmetry of the problem we
can, in principle, choose either branch of the system to estimate $K_{\rho}$.

\section{\label{sec:results}Results}
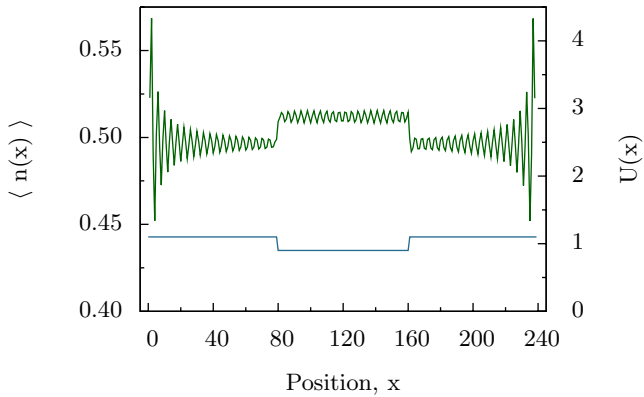
\begin{figure}[h]
  \input{Density}
  \caption{\label{fig:density} Density profile $\langle n(x) \rangle$ for 
    {\it Heterostructure I}, where the on-site Coulomb potential (bottom line on the
    graph with scale on the right) is $U_L=U_R=1.1$ and $U_C=0.9$.  $V(x)=0.0$
    for all sites.  The band filling is $n=0.5$.}
\end{figure}

Using systems with open boundary conditions, finite size effects are induced.
One example of these effects are the local density oscillations and the charge
accumulation close to the edges of the system, shown in
Fig.~\ref{fig:density}.  The charge distribution is expected to be symmetric
around the middle of the chain. We observe, however, that the symmetry is
slightly perturbed, as seen in Fig.~\ref{fig:density}, at the positions where
the Coulomb potential switches values. For our purposes, such small changes
are negligible, specially after taking the average over pairs of correlation
functions, as explained in Sec.~\ref{sec:method}. It is still observed that
the charge density remains fairly homogeneous in the valley of the Coulomb
interaction.

\begin{figure}[h]
  \input{HT020}
  \caption{\label{fig:Hete020} Density-density correlation function for
    {\it Heterostructures I} (continuous line) and {\it II} (broken line). In
    the first case we found $K_{\rho}= 0.864$. In the second case, $K_{\rho}=
    1.093$ only within $R_1$, with $n=0.20$ in both cases.}
\end{figure}
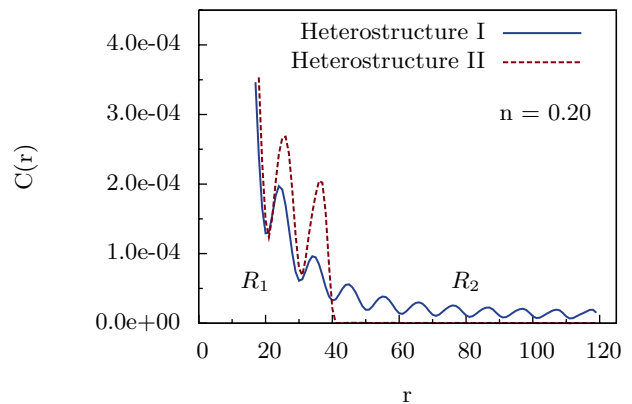

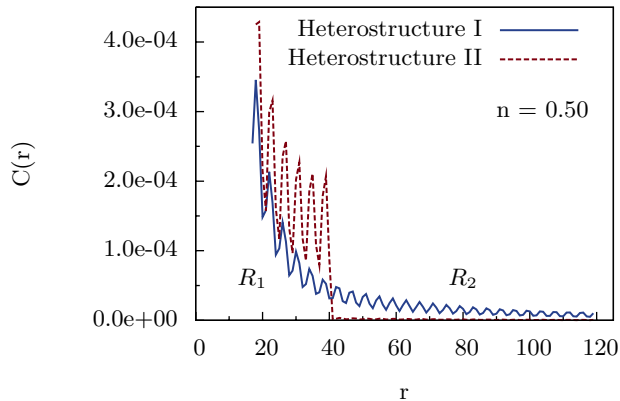
\begin{figure}[h]
  \input{HT050}
  \caption{\label{fig:Hete050} Density-density correlation function for
    {\it Heterostructures I} (continuous line) and {\it II} (broken line). In
    the first case we found $K_{\rho}= 0.838$. In the second case $K_{\rho}=
    1.15$ only within $R_1$, with $n=0.50$ in both cases.}
\end{figure}

To estimate $K_{\rho}$, we fit the values of the numerical data to the leading
term of Eq.~(\ref{correlate}), $K_{\rho}/\pi^2 r^2$, leaving aside, at
least for the moment, the logarithmic corrections. In figures
\ref{fig:Hete020} and \ref{fig:Hete050} the density-density correlations are
shown for both heterostructures and for band fillings of $n=0.20$ and
$n=0.50$. We will refer to the region from the middle of the chain up to the
boundary (at the site $x_R$) of the Coulomb valley as the region $R_1$, and
from this point until the end of the chain as the region $R_2$. In the
following we describe with detail the results for each heterostructure.

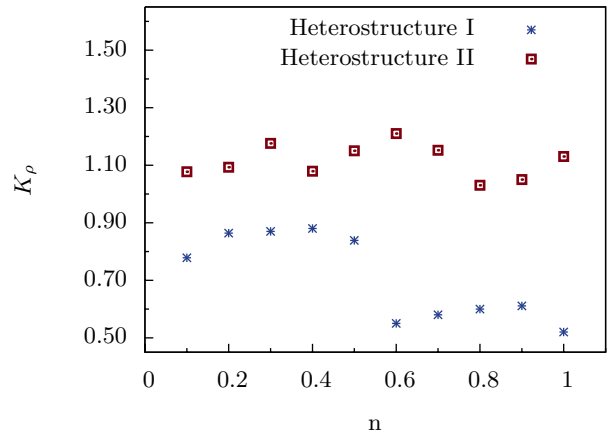
\begin{figure}[h]
  \input{Krho}
  \caption{\label{fig:K_rho} TL parameter $K_\rho$ for both heterostructures as a
    function of the band filling. For {\it Heterostructure II} , only within
    the region $R_1$. For {\it Heterostructure I} $K_\rho < 1.0$ indicating a
    Luttinger liquid behavior. In the second heterostructure $K_\rho > 1.0$,
    which is the benchmark of a Fermi liquid. }
\end{figure}

{\bf Density-density correlation functions for Heterostructure I}.  As a first
structure we take a slightly inhomogeneous Hubbard lattice setting
$U_L=U_R=1.1$ and $U_C=0.9$ with $V(x)=0.0$ for all sites. The valley in the
on-site Coulomb repulsion has sharp edges at the sites $x_L$ and $x_R$, as
shown in the Fig.~\ref{fig:density}. This however, and as seen from the full
line in both Fig.~\ref{fig:Hete020} and in Fig.~\ref{fig:Hete050}, does not
alter significantly the continuous decay of the correlation function. For all
band fillings, $0.10 \le n \le 1.0$, the power-law decay extends beyond the
boundary point and is not completely constrained to any of the regions $R_1$
or $R_2$. In Fig.~\ref{fig:K_rho} the values for the TL parameter are shown as
a function of the band filling. We observe that $K_\rho < 1.0$, which
indicates that spin or charge density waves are present. The $2k_{\rm{F}}$
oscillations can be also observed in the graphs and a closer view is presented
in Fig.~\ref{fig:Friedel}. A fitting of the $2k_{\rm{F}}$ oscillations
succeeded over the whole system only for $n \le 0.5$. In the case of $n > 0.5$,
the fitting of the data was only successful at large distances. This behavior
is reflected on the values of $K_\rho$, as we observe in Fig.~\ref{fig:K_rho}
the two different values sets for the density intervals already mentioned.
With this we confirmed the power-law decay of the correlation functions as it
was possible to determine $K_{\rho}$ also including the first logarithmic
correction.

We compared the results with a similar configuration, this time with an
interaction of the form $U(x)= \cos(\alpha x)$ with $\alpha$ a constant. The
valley around the center of the system remains but the transition on the
potential towards the ends is done in a smoother way. This variation of $U$
resulted in the same values for the correlation functions as already
presented.  showing that the sharp edges of the on-site potential did not
influence strongly the Luttinger liquid behavior of the system.

\begin{figure}[h]
  \input{Oscillations}
  \caption{\label{fig:Friedel} Density-density correlation function
    for {\it Heterostructure I} with $n=0.50$. The crosses show the numerical
    data and the solid line result after fitting including the term
    $A_1\cos(2k_{\rm{F}}x)x^{-(1+K_{\rho})}\mathrm{ln(x)}^{-\frac{3}{2}}$.}
\end{figure}
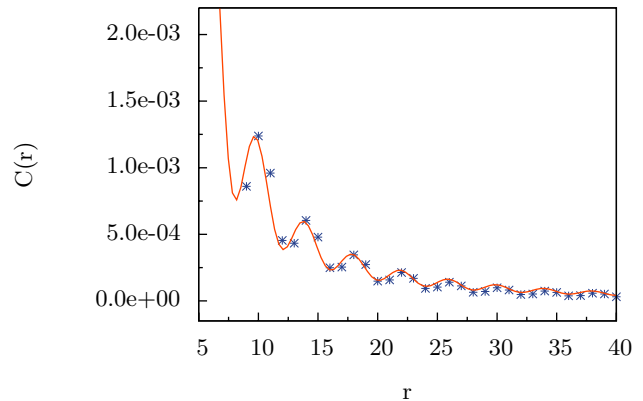

{\bf Density-density correlation functions for Heterostructure II}. In this
case we keep the valley in the Coulomb interaction of the former case:
$U_L=U_R=1.1$ and $U_C=0.9$. Furthermore we simulate two potential walls by
introducing the confining potential $V_{x_L}=V_{x_R}=10.0$ and $V(x)=0.0$ for
the rest of the sites. The results in the case of the {\it Heterostructure II}
distinguished strongly from those previously described.  The introduction of
the confining potential $V(x_L)$ and $V(x_R)$ generated stronger changes from
one region to the other, killing the oscillations beyond the $x_R$ point.  In
Fig.~\ref{fig:Hete020} and Fig.~\ref{fig:Hete050} the broken line shows how
the decay of the correlation function is abruptly interrupted by the
scattering potential, not having further space to fully establish the decay in
the amplitude of the $2k_{\rm{F}}$ oscillations. We found however, that even
in the constrained region $R_1$ the correlation function obeys a power law
decay with $K_\rho > 1.0$, which is indicative of a Fermi liquid behavior.

\section{\label{sec:final}Conclusions}
In this paper we have investigated the behavior of density correlation
functions in one-dimensional heterostructures. We described how junctions
between different types of atoms influence the variation in space of the TL
parameter. The heterostructures as defined can be seen as unions of subunits
with different coupling constants in which the TLL for homogeneous systems is
to be expected.  However, our findings show that a slow variation of the
on-site Coulomb potential, as in the first case, does not interrupt nor split
the decay of the density-density correlation functions between the regions and
the system as a whole behaves as a TLL. Similar systems were investigated
\cite{01_LLS} where the on-site Coulomb potential was turned on and off over
the subchains. For such systems an effective exponent, $K^*_\rho=f(K_{1,\rho},
K_{2,\rho})$, was calculated considering, for example, two subchains which
were assumed as independent, homogeneous TLL's. Using DMRG, their reported
values could only be partially reproduced, namely for densities $n<0.6$.

We have also found a completely different behavior resulting from the
introduction of a scattering potential $V$ at the junctions between the
subunits, as done for the systems in the second case. Our findings in such
case show that the TLL is not a universal feature for one-dimensional systems.
Concerning the dynamics in heterostructures, further work remains to be done.
Transport properties at temperatures different from zero are a key in the
construction of properly tunable electronic devices.

We wish to acknowledge useful discussions with A. Millis and thank C. Kollath
for a helpful revision of the paper.
\bibliography{Literature}
\end{document}

%% file: Density.tex
\begingroup
  \makeatletter
  \providecommand\color[2][]{%
    \GenericError{(gnuplot) \space\space\space\@spaces}{%
      Package color not loaded in conjunction with
      terminal option `colourtext'%
    }{See the gnuplot documentation for explanation.%
    }{Either use 'blacktext' in gnuplot or load the package
      color.sty in LaTeX.}%
    \renewcommand\color[2][]{}%
  }%
  \providecommand\includegraphics[2][]{%
    \GenericError{(gnuplot) \space\space\space\@spaces}{%
      Package graphicx or graphics not loaded%
    }{See the gnuplot documentation for explanation.%
    }{The gnuplot epslatex terminal needs graphicx.sty or graphics.sty.}%
    \renewcommand\includegraphics[2][]{}%
  }%
  \providecommand\rotatebox[2]{#2}%
  \@ifundefined{ifGPcolor}{%
    \newif\ifGPcolor
    \GPcolortrue
  }{}%
  \@ifundefined{ifGPblacktext}{%
    \newif\ifGPblacktext
    \GPblacktexttrue
  }{}%
  \let\gplgaddtomacro\g@addto@macro
  \gdef\gplbacktext{}%
  \gdef\gplfronttext{}%
  \makeatother
  \ifGPblacktext
    \def\colorrgb#1{}%
    \def\colorgray#1{}%
  \else
    \ifGPcolor
      \def\colorrgb#1{\color[rgb]{#1}}%
      \def\colorgray#1{\color[gray]{#1}}%
      \expandafter\def\csname LTw\endcsname{\color{white}}%
      \expandafter\def\csname LTb\endcsname{\color{black}}%
      \expandafter\def\csname LTa\endcsname{\color{black}}%
      \expandafter\def\csname LT0\endcsname{\color[rgb]{1,0,0}}%
      \expandafter\def\csname LT1\endcsname{\color[rgb]{0,1,0}}%
      \expandafter\def\csname LT2\endcsname{\color[rgb]{0,0,1}}%
      \expandafter\def\csname LT3\endcsname{\color[rgb]{1,0,1}}%
      \expandafter\def\csname LT4\endcsname{\color[rgb]{0,1,1}}%
      \expandafter\def\csname LT5\endcsname{\color[rgb]{1,1,0}}%
      \expandafter\def\csname LT6\endcsname{\color[rgb]{0,0,0}}%
      \expandafter\def\csname LT7\endcsname{\color[rgb]{1,0.3,0}}%
      \expandafter\def\csname LT8\endcsname{\color[rgb]{0.5,0.5,0.5}}%
    \else
      \def\colorrgb#1{\color{black}}%
      \def\colorgray#1{\color[gray]{#1}}%
      \expandafter\def\csname LTw\endcsname{\color{white}}%
      \expandafter\def\csname LTb\endcsname{\color{black}}%
      \expandafter\def\csname LTa\endcsname{\color{black}}%
      \expandafter\def\csname LT0\endcsname{\color{black}}%
      \expandafter\def\csname LT1\endcsname{\color{black}}%
      \expandafter\def\csname LT2\endcsname{\color{black}}%
      \expandafter\def\csname LT3\endcsname{\color{black}}%
      \expandafter\def\csname LT4\endcsname{\color{black}}%
      \expandafter\def\csname LT5\endcsname{\color{black}}%
      \expandafter\def\csname LT6\endcsname{\color{black}}%
      \expandafter\def\csname LT7\endcsname{\color{black}}%
      \expandafter\def\csname LT8\endcsname{\color{black}}%
    \fi
  \fi
  \setlength{\unitlength}{0.0500bp}%
  \begin{picture}(5040.00,3528.00)%
    \gplgaddtomacro\gplbacktext{%
      \csname LTb\endcsname%
      \put(990,814){\makebox(0,0)[r]{\strut{}0.40}}%
      \put(990,1470){\makebox(0,0)[r]{\strut{}0.45}}%
      \put(990,2125){\makebox(0,0)[r]{\strut{}0.50}}%
      \put(990,2781){\makebox(0,0)[r]{\strut{}0.55}}%
      \put(1183,594){\makebox(0,0){\strut{} 0}}%
      \put(1673,594){\makebox(0,0){\strut{} 40}}%
      \put(2162,594){\makebox(0,0){\strut{} 80}}%
      \put(2652,594){\makebox(0,0){\strut{} 120}}%
      \put(3142,594){\makebox(0,0){\strut{} 160}}%
      \put(3631,594){\makebox(0,0){\strut{} 200}}%
      \put(4121,594){\makebox(0,0){\strut{} 240}}%
      \put(4314,814){\makebox(0,0)[l]{\strut{} 0}}%
      \put(4314,1324){\makebox(0,0)[l]{\strut{} 1}}%
      \put(4314,1834){\makebox(0,0)[l]{\strut{} 2}}%
      \put(4314,2344){\makebox(0,0)[l]{\strut{} 3}}%
      \put(4314,2854){\makebox(0,0)[l]{\strut{} 4}}%
      \put(220,1961){\rotatebox{90}{\makebox(0,0){$\langle$ n(x) $\rangle$}}}%
      \put(4819,1961){\rotatebox{90}{\makebox(0,0){\strut{}U(x)}}}%
      \put(2652,264){\makebox(0,0){\strut{}Position, x}}%
    }%
    \gplgaddtomacro\gplfronttext{%
    }%
    \gplbacktext
    \put(0,0){\includegraphics{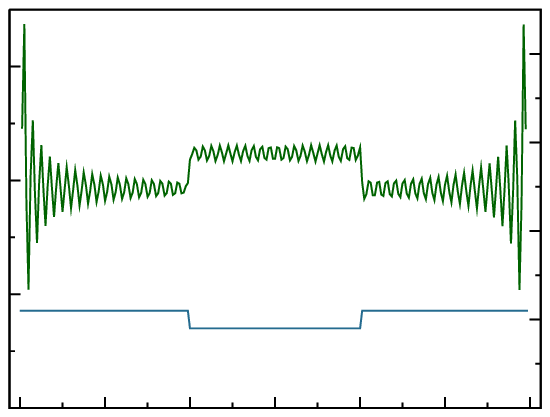}}%
    \gplfronttext
  \end{picture}%
\endgroup


%% file: Oscillations.tex
\begingroup
  \makeatletter
  \providecommand\color[2][]{%
    \GenericError{(gnuplot) \space\space\space\@spaces}{%
      Package color not loaded in conjunction with
      terminal option `colourtext'%
    }{See the gnuplot documentation for explanation.%
    }{Either use 'blacktext' in gnuplot or load the package
      color.sty in LaTeX.}%
    \renewcommand\color[2][]{}%
  }%
  \providecommand\includegraphics[2][]{%
    \GenericError{(gnuplot) \space\space\space\@spaces}{%
      Package graphicx or graphics not loaded%
    }{See the gnuplot documentation for explanation.%
    }{The gnuplot epslatex terminal needs graphicx.sty or graphics.sty.}%
    \renewcommand\includegraphics[2][]{}%
  }%
  \providecommand\rotatebox[2]{#2}%
  \@ifundefined{ifGPcolor}{%
    \newif\ifGPcolor
    \GPcolortrue
  }{}%
  \@ifundefined{ifGPblacktext}{%
    \newif\ifGPblacktext
    \GPblacktexttrue
  }{}%
  \let\gplgaddtomacro\g@addto@macro
  \gdef\gplbacktext{}%
  \gdef\gplfronttext{}%
  \makeatother
  \ifGPblacktext
    \def\colorrgb#1{}%
    \def\colorgray#1{}%
  \else
    \ifGPcolor
      \def\colorrgb#1{\color[rgb]{#1}}%
      \def\colorgray#1{\color[gray]{#1}}%
      \expandafter\def\csname LTw\endcsname{\color{white}}%
      \expandafter\def\csname LTb\endcsname{\color{black}}%
      \expandafter\def\csname LTa\endcsname{\color{black}}%
      \expandafter\def\csname LT0\endcsname{\color[rgb]{1,0,0}}%
      \expandafter\def\csname LT1\endcsname{\color[rgb]{0,1,0}}%
      \expandafter\def\csname LT2\endcsname{\color[rgb]{0,0,1}}%
      \expandafter\def\csname LT3\endcsname{\color[rgb]{1,0,1}}%
      \expandafter\def\csname LT4\endcsname{\color[rgb]{0,1,1}}%
      \expandafter\def\csname LT5\endcsname{\color[rgb]{1,1,0}}%
      \expandafter\def\csname LT6\endcsname{\color[rgb]{0,0,0}}%
      \expandafter\def\csname LT7\endcsname{\color[rgb]{1,0.3,0}}%
      \expandafter\def\csname LT8\endcsname{\color[rgb]{0.5,0.5,0.5}}%
    \else
      \def\colorrgb#1{\color{black}}%
      \def\colorgray#1{\color[gray]{#1}}%
      \expandafter\def\csname LTw\endcsname{\color{white}}%
      \expandafter\def\csname LTb\endcsname{\color{black}}%
      \expandafter\def\csname LTa\endcsname{\color{black}}%
      \expandafter\def\csname LT0\endcsname{\color{black}}%
      \expandafter\def\csname LT1\endcsname{\color{black}}%
      \expandafter\def\csname LT2\endcsname{\color{black}}%
      \expandafter\def\csname LT3\endcsname{\color{black}}%
      \expandafter\def\csname LT4\endcsname{\color{black}}%
      \expandafter\def\csname LT5\endcsname{\color{black}}%
      \expandafter\def\csname LT6\endcsname{\color{black}}%
      \expandafter\def\csname LT7\endcsname{\color{black}}%
      \expandafter\def\csname LT8\endcsname{\color{black}}%
    \fi
  \fi
  \setlength{\unitlength}{0.0500bp}%
  \begin{picture}(5040.00,3528.00)%
    \gplgaddtomacro\gplbacktext{%
      \csname LTb\endcsname%
      \put(1386,932){\makebox(0,0)[r]{\strut{}0.0e+00}}%
      \put(1386,1434){\makebox(0,0)[r]{\strut{}5.0e-04}}%
      \put(1386,1936){\makebox(0,0)[r]{\strut{}1.0e-03}}%
      \put(1386,2439){\makebox(0,0)[r]{\strut{}1.5e-03}}%
      \put(1386,2941){\makebox(0,0)[r]{\strut{}2.0e-03}}%
      \put(1518,561){\makebox(0,0){\strut{} 5}}%
      \put(1968,561){\makebox(0,0){\strut{} 10}}%
      \put(2417,561){\makebox(0,0){\strut{} 15}}%
      \put(2867,561){\makebox(0,0){\strut{} 20}}%
      \put(3317,561){\makebox(0,0){\strut{} 25}}%
      \put(3767,561){\makebox(0,0){\strut{} 30}}%
      \put(4216,561){\makebox(0,0){\strut{} 35}}%
      \put(4666,561){\makebox(0,0){\strut{} 40}}%
      \put(220,1961){\rotatebox{90}{\makebox(0,0){\strut{}C(r)}}}%
      \put(3092,231){\makebox(0,0){\strut{}r}}%
    }%
    \gplgaddtomacro\gplfronttext{%
    }%
    \gplbacktext
    \put(0,0){\includegraphics{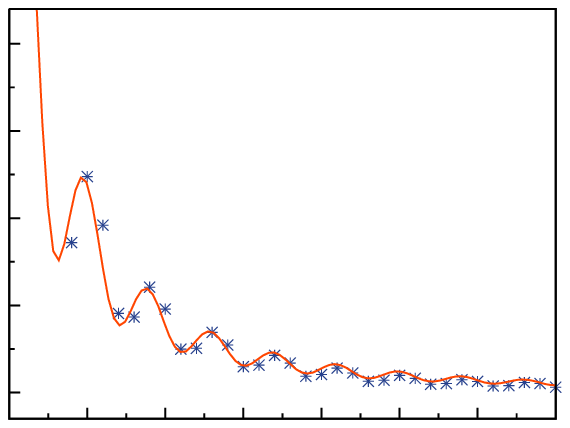}}%
    \gplfronttext
  \end{picture}%
\endgroup


%% file: HT020.tex
\begingroup
  \makeatletter
  \providecommand\color[2][]{%
    \GenericError{(gnuplot) \space\space\space\@spaces}{%
      Package color not loaded in conjunction with
      terminal option `colourtext'%
    }{See the gnuplot documentation for explanation.%
    }{Either use 'blacktext' in gnuplot or load the package
      color.sty in LaTeX.}%
    \renewcommand\color[2][]{}%
  }%
  \providecommand\includegraphics[2][]{%
    \GenericError{(gnuplot) \space\space\space\@spaces}{%
      Package graphicx or graphics not loaded%
    }{See the gnuplot documentation for explanation.%
    }{The gnuplot epslatex terminal needs graphicx.sty or graphics.sty.}%
    \renewcommand\includegraphics[2][]{}%
  }%
  \providecommand\rotatebox[2]{#2}%
  \@ifundefined{ifGPcolor}{%
    \newif\ifGPcolor
    \GPcolortrue
  }{}%
  \@ifundefined{ifGPblacktext}{%
    \newif\ifGPblacktext
    \GPblacktexttrue
  }{}%
  \let\gplgaddtomacro\g@addto@macro
  \gdef\gplbacktext{}%
  \gdef\gplfronttext{}%
  \makeatother
  \ifGPblacktext
    \def\colorrgb#1{}%
    \def\colorgray#1{}%
  \else
    \ifGPcolor
      \def\colorrgb#1{\color[rgb]{#1}}%
      \def\colorgray#1{\color[gray]{#1}}%
      \expandafter\def\csname LTw\endcsname{\color{white}}%
      \expandafter\def\csname LTb\endcsname{\color{black}}%
      \expandafter\def\csname LTa\endcsname{\color{black}}%
      \expandafter\def\csname LT0\endcsname{\color[rgb]{1,0,0}}%
      \expandafter\def\csname LT1\endcsname{\color[rgb]{0,1,0}}%
      \expandafter\def\csname LT2\endcsname{\color[rgb]{0,0,1}}%
      \expandafter\def\csname LT3\endcsname{\color[rgb]{1,0,1}}%
      \expandafter\def\csname LT4\endcsname{\color[rgb]{0,1,1}}%
      \expandafter\def\csname LT5\endcsname{\color[rgb]{1,1,0}}%
      \expandafter\def\csname LT6\endcsname{\color[rgb]{0,0,0}}%
      \expandafter\def\csname LT7\endcsname{\color[rgb]{1,0.3,0}}%
      \expandafter\def\csname LT8\endcsname{\color[rgb]{0.5,0.5,0.5}}%
    \else
      \def\colorrgb#1{\color{black}}%
      \def\colorgray#1{\color[gray]{#1}}%
      \expandafter\def\csname LTw\endcsname{\color{white}}%
      \expandafter\def\csname LTb\endcsname{\color{black}}%
      \expandafter\def\csname LTa\endcsname{\color{black}}%
      \expandafter\def\csname LT0\endcsname{\color{black}}%
      \expandafter\def\csname LT1\endcsname{\color{black}}%
      \expandafter\def\csname LT2\endcsname{\color{black}}%
      \expandafter\def\csname LT3\endcsname{\color{black}}%
      \expandafter\def\csname LT4\endcsname{\color{black}}%
      \expandafter\def\csname LT5\endcsname{\color{black}}%
      \expandafter\def\csname LT6\endcsname{\color{black}}%
      \expandafter\def\csname LT7\endcsname{\color{black}}%
      \expandafter\def\csname LT8\endcsname{\color{black}}%
    \fi
  \fi
  \setlength{\unitlength}{0.0500bp}%
  \begin{picture}(5040.00,3528.00)%
    \gplgaddtomacro\gplbacktext{%
      \csname LTb\endcsname%
      \put(1386,781){\makebox(0,0)[r]{\strut{}0.0e+00}}%
      \put(1386,1306){\makebox(0,0)[r]{\strut{}1.0e-04}}%
      \put(1386,1830){\makebox(0,0)[r]{\strut{}2.0e-04}}%
      \put(1386,2355){\makebox(0,0)[r]{\strut{}3.0e-04}}%
      \put(1386,2880){\makebox(0,0)[r]{\strut{}4.0e-04}}%
      \put(1518,561){\makebox(0,0){\strut{} 0}}%
      \put(2022,561){\makebox(0,0){\strut{} 20}}%
      \put(2525,561){\makebox(0,0){\strut{} 40}}%
      \put(3029,561){\makebox(0,0){\strut{} 60}}%
      \put(3533,561){\makebox(0,0){\strut{} 80}}%
      \put(4036,561){\makebox(0,0){\strut{} 100}}%
      \put(4540,561){\makebox(0,0){\strut{} 120}}%
      \put(220,1961){\rotatebox{90}{\makebox(0,0){\strut{}C(r)}}}%
      \put(3092,231){\makebox(0,0){\strut{}r}}%
      \put(3785,2355){\makebox(0,0)[l]{\strut{}n = 0.20}}%
      \put(1946,1096){\makebox(0,0){$R_1$}}%
      \put(3533,1096){\makebox(0,0){$R_2$}}%
    }%
    \gplgaddtomacro\gplfronttext{%
      \csname LTb\endcsname%
      \put(3679,2969){\makebox(0,0)[r]{\strut{}Heterostructure I}}%
      \csname LTb\endcsname%
      \put(3679,2749){\makebox(0,0)[r]{\strut{}Heterostructure II}}%
    }%
    \gplbacktext
    \put(0,0){\includegraphics{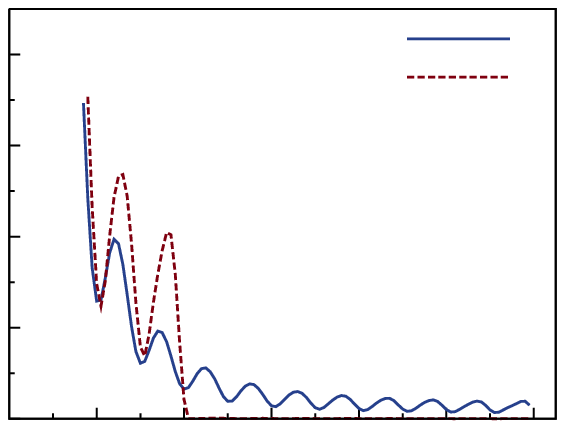}}%
    \gplfronttext
  \end{picture}%
\endgroup


%% file: HT050.tex
\begingroup
  \makeatletter
  \providecommand\color[2][]{%
    \GenericError{(gnuplot) \space\space\space\@spaces}{%
      Package color not loaded in conjunction with
      terminal option `colourtext'%
    }{See the gnuplot documentation for explanation.%
    }{Either use 'blacktext' in gnuplot or load the package
      color.sty in LaTeX.}%
    \renewcommand\color[2][]{}%
  }%
  \providecommand\includegraphics[2][]{%
    \GenericError{(gnuplot) \space\space\space\@spaces}{%
      Package graphicx or graphics not loaded%
    }{See the gnuplot documentation for explanation.%
    }{The gnuplot epslatex terminal needs graphicx.sty or graphics.sty.}%
    \renewcommand\includegraphics[2][]{}%
  }%
  \providecommand\rotatebox[2]{#2}%
  \@ifundefined{ifGPcolor}{%
    \newif\ifGPcolor
    \GPcolortrue
  }{}%
  \@ifundefined{ifGPblacktext}{%
    \newif\ifGPblacktext
    \GPblacktexttrue
  }{}%
  \let\gplgaddtomacro\g@addto@macro
  \gdef\gplbacktext{}%
  \gdef\gplfronttext{}%
  \makeatother
  \ifGPblacktext
    \def\colorrgb#1{}%
    \def\colorgray#1{}%
  \else
    \ifGPcolor
      \def\colorrgb#1{\color[rgb]{#1}}%
      \def\colorgray#1{\color[gray]{#1}}%
      \expandafter\def\csname LTw\endcsname{\color{white}}%
      \expandafter\def\csname LTb\endcsname{\color{black}}%
      \expandafter\def\csname LTa\endcsname{\color{black}}%
      \expandafter\def\csname LT0\endcsname{\color[rgb]{1,0,0}}%
      \expandafter\def\csname LT1\endcsname{\color[rgb]{0,1,0}}%
      \expandafter\def\csname LT2\endcsname{\color[rgb]{0,0,1}}%
      \expandafter\def\csname LT3\endcsname{\color[rgb]{1,0,1}}%
      \expandafter\def\csname LT4\endcsname{\color[rgb]{0,1,1}}%
      \expandafter\def\csname LT5\endcsname{\color[rgb]{1,1,0}}%
      \expandafter\def\csname LT6\endcsname{\color[rgb]{0,0,0}}%
      \expandafter\def\csname LT7\endcsname{\color[rgb]{1,0.3,0}}%
      \expandafter\def\csname LT8\endcsname{\color[rgb]{0.5,0.5,0.5}}%
    \else
      \def\colorrgb#1{\color{black}}%
      \def\colorgray#1{\color[gray]{#1}}%
      \expandafter\def\csname LTw\endcsname{\color{white}}%
      \expandafter\def\csname LTb\endcsname{\color{black}}%
      \expandafter\def\csname LTa\endcsname{\color{black}}%
      \expandafter\def\csname LT0\endcsname{\color{black}}%
      \expandafter\def\csname LT1\endcsname{\color{black}}%
      \expandafter\def\csname LT2\endcsname{\color{black}}%
      \expandafter\def\csname LT3\endcsname{\color{black}}%
      \expandafter\def\csname LT4\endcsname{\color{black}}%
      \expandafter\def\csname LT5\endcsname{\color{black}}%
      \expandafter\def\csname LT6\endcsname{\color{black}}%
      \expandafter\def\csname LT7\endcsname{\color{black}}%
      \expandafter\def\csname LT8\endcsname{\color{black}}%
    \fi
  \fi
  \setlength{\unitlength}{0.0500bp}%
  \begin{picture}(5040.00,3528.00)%
    \gplgaddtomacro\gplbacktext{%
      \csname LTb\endcsname%
      \put(1386,781){\makebox(0,0)[r]{\strut{}0.0e+00}}%
      \put(1386,1306){\makebox(0,0)[r]{\strut{}1.0e-04}}%
      \put(1386,1830){\makebox(0,0)[r]{\strut{}2.0e-04}}%
      \put(1386,2355){\makebox(0,0)[r]{\strut{}3.0e-04}}%
      \put(1386,2880){\makebox(0,0)[r]{\strut{}4.0e-04}}%
      \put(1518,561){\makebox(0,0){\strut{} 0}}%
      \put(2022,561){\makebox(0,0){\strut{} 20}}%
      \put(2525,561){\makebox(0,0){\strut{} 40}}%
      \put(3029,561){\makebox(0,0){\strut{} 60}}%
      \put(3533,561){\makebox(0,0){\strut{} 80}}%
      \put(4036,561){\makebox(0,0){\strut{} 100}}%
      \put(4540,561){\makebox(0,0){\strut{} 120}}%
      \put(220,1961){\rotatebox{90}{\makebox(0,0){\strut{}C(r)}}}%
      \put(3092,231){\makebox(0,0){\strut{}r}}%
      \put(3785,2355){\makebox(0,0)[l]{\strut{}n = 0.50}}%
      \put(1946,1096){\makebox(0,0){$R_1$}}%
      \put(3533,1096){\makebox(0,0){$R_2$}}%
    }%
    \gplgaddtomacro\gplfronttext{%
      \csname LTb\endcsname%
      \put(3679,2969){\makebox(0,0)[r]{\strut{}Heterostructure I}}%
      \csname LTb\endcsname%
      \put(3679,2749){\makebox(0,0)[r]{\strut{}Heterostructure II}}%
    }%
    \gplbacktext
    \put(0,0){\includegraphics{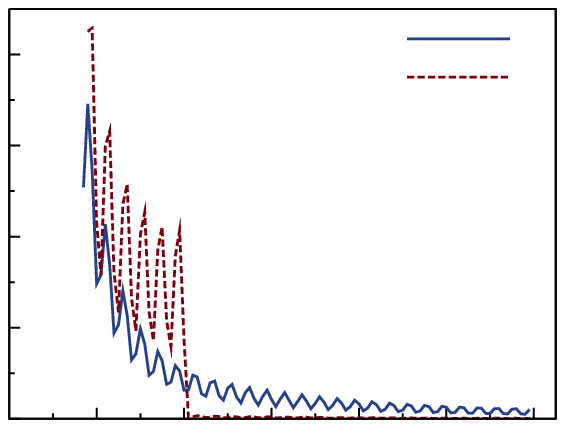}}%
    \gplfronttext
  \end{picture}%
\endgroup


%% file: Krho.tex
\begingroup
  \makeatletter
  \providecommand\color[2][]{%
    \GenericError{(gnuplot) \space\space\space\@spaces}{%
      Package color not loaded in conjunction with
      terminal option `colourtext'%
    }{See the gnuplot documentation for explanation.%
    }{Either use 'blacktext' in gnuplot or load the package
      color.sty in LaTeX.}%
    \renewcommand\color[2][]{}%
  }%
  \providecommand\includegraphics[2][]{%
    \GenericError{(gnuplot) \space\space\space\@spaces}{%
      Package graphicx or graphics not loaded%
    }{See the gnuplot documentation for explanation.%
    }{The gnuplot epslatex terminal needs graphicx.sty or graphics.sty.}%
    \renewcommand\includegraphics[2][]{}%
  }%
  \providecommand\rotatebox[2]{#2}%
  \@ifundefined{ifGPcolor}{%
    \newif\ifGPcolor
    \GPcolortrue
  }{}%
  \@ifundefined{ifGPblacktext}{%
    \newif\ifGPblacktext
    \GPblacktexttrue
  }{}%
  \let\gplgaddtomacro\g@addto@macro
  \gdef\gplbacktext{}%
  \gdef\gplfronttext{}%
  \makeatother
  \ifGPblacktext
    \def\colorrgb#1{}%
    \def\colorgray#1{}%
  \else
    \ifGPcolor
      \def\colorrgb#1{\color[rgb]{#1}}%
      \def\colorgray#1{\color[gray]{#1}}%
      \expandafter\def\csname LTw\endcsname{\color{white}}%
      \expandafter\def\csname LTb\endcsname{\color{black}}%
      \expandafter\def\csname LTa\endcsname{\color{black}}%
      \expandafter\def\csname LT0\endcsname{\color[rgb]{1,0,0}}%
      \expandafter\def\csname LT1\endcsname{\color[rgb]{0,1,0}}%
      \expandafter\def\csname LT2\endcsname{\color[rgb]{0,0,1}}%
      \expandafter\def\csname LT3\endcsname{\color[rgb]{1,0,1}}%
      \expandafter\def\csname LT4\endcsname{\color[rgb]{0,1,1}}%
      \expandafter\def\csname LT5\endcsname{\color[rgb]{1,1,0}}%
      \expandafter\def\csname LT6\endcsname{\color[rgb]{0,0,0}}%
      \expandafter\def\csname LT7\endcsname{\color[rgb]{1,0.3,0}}%
      \expandafter\def\csname LT8\endcsname{\color[rgb]{0.5,0.5,0.5}}%
    \else
      \def\colorrgb#1{\color{black}}%
      \def\colorgray#1{\color[gray]{#1}}%
      \expandafter\def\csname LTw\endcsname{\color{white}}%
      \expandafter\def\csname LTb\endcsname{\color{black}}%
      \expandafter\def\csname LTa\endcsname{\color{black}}%
      \expandafter\def\csname LT0\endcsname{\color{black}}%
      \expandafter\def\csname LT1\endcsname{\color{black}}%
      \expandafter\def\csname LT2\endcsname{\color{black}}%
      \expandafter\def\csname LT3\endcsname{\color{black}}%
      \expandafter\def\csname LT4\endcsname{\color{black}}%
      \expandafter\def\csname LT5\endcsname{\color{black}}%
      \expandafter\def\csname LT6\endcsname{\color{black}}%
      \expandafter\def\csname LT7\endcsname{\color{black}}%
      \expandafter\def\csname LT8\endcsname{\color{black}}%
    \fi
  \fi
  \setlength{\unitlength}{0.0500bp}%
  \begin{picture}(5040.00,3528.00)%
    \gplgaddtomacro\gplbacktext{%
      \csname LTb\endcsname%
      \put(1026,768){\makebox(0,0)[r]{\strut{}0.50}}%
      \put(1026,1202){\makebox(0,0)[r]{\strut{}0.70}}%
      \put(1026,1636){\makebox(0,0)[r]{\strut{}0.90}}%
      \put(1026,2070){\makebox(0,0)[r]{\strut{}1.10}}%
      \put(1026,2504){\makebox(0,0)[r]{\strut{}1.30}}%
      \put(1026,2938){\makebox(0,0)[r]{\strut{}1.50}}%
      \put(1158,440){\makebox(0,0){\strut{} 0}}%
      \put(1789,440){\makebox(0,0){\strut{} 0.2}}%
      \put(2421,440){\makebox(0,0){\strut{} 0.4}}%
      \put(3052,440){\makebox(0,0){\strut{} 0.6}}%
      \put(3683,440){\makebox(0,0){\strut{} 0.8}}%
      \put(4314,440){\makebox(0,0){\strut{} 1}}%
      \put(256,1962){\rotatebox{90}{\makebox(0,0){\strut{}$K_\rho$}}}%
      \put(2894,110){\makebox(0,0){\strut{}n}}%
    }%
    \gplgaddtomacro\gplfronttext{%
      \csname LTb\endcsname%
      \put(3643,3091){\makebox(0,0)[r]{\strut{}Heterostructure I}}%
      \csname LTb\endcsname%
      \put(3643,2871){\makebox(0,0)[r]{\strut{}Heterostructure II}}%
    }%
    \gplbacktext
    \put(0,0){\includegraphics{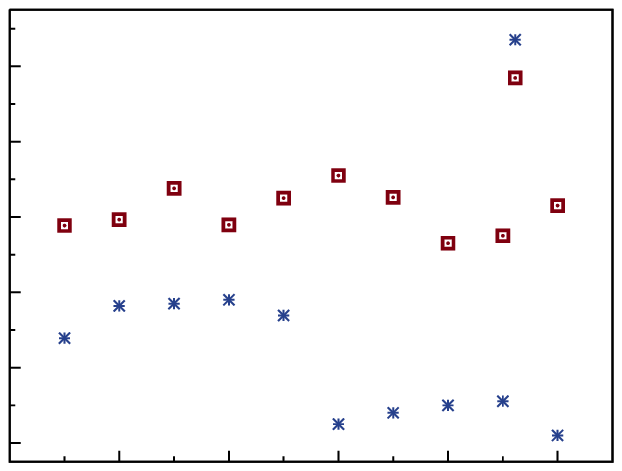}}%
    \gplfronttext
  \end{picture}%
\endgroup
